\documentclass{llncs}
\usepackage{graphics}

\begin{document}

\title{The efficient generation of unstructured control volumes in 2D and 3D}

\author{Jacek Leszczynski, Sebastian Pluta}

\institute{Technical University of Czestochowa, \\ 
           Institute of Mathematics and Computer Science, \\
           ul. Dabrowskiego 73, 42-200 Czestochowa, Poland \\
           \email{jale@k2.pcz.czest.pl}, \email{pluta@matinf.pcz.czest.pl}}

\titlerunning{The efficient generation ...}
\authorrunning{Jacek Leszczynski et al}
\maketitle              % typeset the title of the contribution

\begin{abstract}
 Many problems in engineering, chemistry and physics require the representation
 of solutions in complex geometries. In the paper we deal with a~problem of
 unstructured mesh generation for the control volume method. We propose an
 algorithm which bases on the spheres generation in central points of the control
 volumes.
\end{abstract}

%\noindent {\bf Key words:} unstructured mesh, control volume, Delaunay
%                           triangulation

\section{Introduction}
An unstructured mesh generator can solve many of the problems associated with
the structured meshes. The unstructured meshes are suitable for complex
geometries, especially when we use the control volume method \cite{Patankar}. The
control volume method on the one hand, is based upon the simple physical principle
of a flux balance. Nevertheless, the generated meshes have to fulfill some
restrictions for such method. One of the conditions says than lines connecting the
central point in the control volume with neighbour central points have to be
perpendicular to the sides of the neighbour volumes. However, the neighbourhood
of each neighbour or cell in an unstructured mesh must be defined explicitly. This
is one disadvantage of unstructured meshes because it has to reserve large storage
of the computer memory. Nevertheless, the advantages like easy for handling
adaptability in time, ability to generate meshes about arbitrary geometries.
We can divide such meshes into the cell-vertex control volumes and the
cell-centered control volumes. In the cell-vertex volumes, all elements containing
the relevant point are applied as a~control volume, so the control volumes overlap.
In the cell-centered control volumes the elements are subdivided and the control
volumes do not overlap.

In this paper we propose a~generator for unstructured cell-centered volumes,
which is useful for the control volume method \cite{Patankar}. The mesh is
constructed for a~number of points randomly located in a~domain and the domain
border.

\section{Mathematical background}
We turn our attention in two and three dimensional space. Let us consider
a~domain \(\Omega \subset R^3\) with smooth its boundary \(\partial\Omega\). In
such domain and  boundary we generate randomly a~set of points \(P_i(a,b,c)\),
where \(i=1..N\). The points establish cell-centers, around of which we determine
convex polygons. The polygons are the non-overlapping control volumes which are
applied in the control volume method \cite{Patankar}. We can formulate two groups
of basic assumptions necessary for the meshes construction as follows:
\begin{enumerate}
 \item global conditions:
  \subitem such domain has to be consistent,
  \subitem edges of some polygon have the same lengths to the corresponding edges
           of neighbour polygons,
 \item local conditions:
  \subitem inside a polygon one may found only one random point called the
           cell-center or the point,
  \subitem an edge of the polygon has to be perpendicular to the line
           connecting two points.
\end{enumerate}
If we fulfill the global assumptions, we may discretize the domain \(\Omega\)
through its division into the triangles in \(R^2\) space or into the tetrahedrons
in \(R^3\) space. Such discretization is a~standard solution, which one may find
in convex geometry \cite{Handbook} through Delaunay
triangulation~\cite{Chiang,Orkisz,Renka}. Regarding to the global and local
conditions we take into consideration a~fact, that we cannot generate polygons in
\(R^2\) which overlap neighbour polygons. The fact also exist for the generated
tetrahedrons in \(R^3\) space.

In this paper we propose a~novel algorithm for generation of control volumes. We
start the algorithm in two dimensions because one can see how it works. We
consider circles with unknown radiuses which centers are located in the points,
which are randomly located in the domain \(\Omega\). Let us assume, that within
the triangle created by the points the circles only in the one point crosses. The
point is a~vertice of control volumes generated around the point. Following that
we need to solve the system of equations
\begin{equation} \label{eq1}
 (\widehat{x}-a_{l})^{2}+(\widehat{y}-b_{l})^{2}=r_{l}^{2},   (l=1,2,3).
\end{equation}
Assuming, that the radiuses \(r_l\) of the circles are unknown, we obtain
a~point \(Q\big(\widehat{x}(r_1,r_2,r_3),\widehat{y}(r_1,r_2,r_3)\big)\) which
coordinates are solution of the system (\ref{eq1})
\begin{equation} \label{eq2}
 \left.
 \begin{array}{c}
  \widehat{x}(r_{1},\, r_{2},\, r_{3})=\frac{-(b_{2}-b_{3})\cdot
   r_{1}^{2}+(b_{1}-b_{3})\cdot r_{2}^{2}-(b_{1}-b_{2})\cdot r_{3}^{2}+
   (a_{1}^{2}+b_{1}^{2})\cdot (b_{2}-b_{3})}{2\cdot\left[(a_{1}-a_{2})\cdot
   (b_{1}-b_{3})-(a_{1}-a_{3})\cdot (b_{1}-b_{2})\right] }+\\
   -\frac{(a_{2}^{2}+b_{2}^{2})\cdot (b_{1}-b_{3})+(a_{3}^{2}+b_{3}^{2})\cdot
   (b_{1}-b_{2})}{2\cdot \left[(a_{1}-a_{2})\cdot (b_{1}-b_{3})-(a_{1}-a_{3})\cdot
   (b_{1}-b_{2})\right] }\\
  \widehat{y}(r_{1},\, r_{2},\, r_{3})=\frac{(a_{2}-a_{3})\cdot
   r_{1}^{2}-(a_{1}-a_{3})\cdot r_{2}^{2}+(a_{1}-a_{2})\cdot
   r_{3}^{2}-(a_{1}^{2}+b_{1}^{2})\cdot (a_{2}-a_{3})}{2\cdot
   \left[ (a_{1}-a_{2})\cdot (b_{1}-b_{3})-(a_{1}-a_{3})\cdot
   (b_{1}-b_{2})\right] }+\\
   +\frac{(a_{2}^{2}+b_{2}^{2})\cdot (a_{1}-a_{3})-(a_{3}^{2}+b_{3}^{2})\cdot
   (a_{1}-a_{2})}{2\cdot \left[ (a_{1}-a_{2})\cdot (b_{1}-b_{3})-(a_{1}-a_{3})\cdot
   (b_{1}-b_{2})\right] }
 \end{array}\right. 
\end{equation}
The indexes in formulas (\ref{eq1}) and (\ref{eq2}) correspond to the local case.

In the global case we have random points \(P_i(a,b,c),   i=1..N\) which are
generated inside the domain \(\Omega\) and its boundary \(\partial\Omega\). We
use Delaunay triangulation~\cite{Chiang,Orkisz,Renka} that to find neighbourhood
of point \(P_i\) defined by several points
\(P_{j(i,1)}\dots P_{j(i,k)}\dots P_{j(i,M(i))}\). The index \(j(i,k)\) is
a~function which establishes a~point number having neighbourhood with the
\(i\)-th point. The temporal index \(k\) varies from \(1\) to \(M(i)\). The
\(M(i)\) is a~number of points corresponding to the \(i\)-th point. Three points
\(P_i, P_{j(i,k)}, P_{j(i,k+1)}\) define a~triangle. When the temporal index
\(k+1\) exceeds \(M(i)\) then we have \(P_{j(i,k+1)}=P_{j(i,1)}\). Moreover, we
take into consideration a~condition \(j(i,k)\neq i\). We extend a~definition of
formula~(\ref{eq2}) for the global case putting \((a_i,b_i)\),
\((a_{j(i,k)},b_{j(i,k)})\), \((a_{j(i,k+1)},b_{j(i,k+1)})\) and
\((r_i,r_{j(i,k)},r_{j(i,k+1)})\) respectively into variables with established
indexes. In our approach, we are looking for the radiuses \(r_i\). Taking into
consideration of such fact we formulate a~range of radiuses variation as
\begin{equation} \label{eq3}
 \max\limits_{k} \left[(L_{i,j(i,k)}-r_{max\,j(i,k)})\right]<r_{i}<r_{max\,i},
\end{equation}
where \(L_{i,j(i,k)}\) is a~distance between two neighbour points defined as
\begin{equation} \label{eq4}
 L_{i,j(i)}=\sqrt{(a_{i}-a_{j(i)})^{2}+(b_{i}-b_{j(i)})^{2}}. 
\end{equation}
Maximal radius of a~circle is formed as
\begin{equation} \label{eq41}
 r_{max\,i}=\min\limits_k\left[h_{i,j(i,k)}\right],
\end{equation}
where \(h_{i,j(i,k)}\) is a~distance defined in acute triangles as
\begin{equation} \label{eq5}
 \begin{array}{c}
  h_{i,j(i,k)}=\frac{\vert(b_{j(i,k)}-b_{j(i,k+1)})a_{i}-(a_{j(i,k)}-
  a_{j(i,k+1)})b_{i}\vert}{\sqrt{(b_{j(i,k)}-b_{j(i,k+1)})^2+(a_{j(i,k)}-
  a_{j(i,k+1)})^2}}+\\
  +\frac{\vert(a_{j(i,k)}-a_{j(i,k+1)})b_{j(i,k+1)}-(b_{j(i,k)}-b_{j(i,k+1)})
  a_{j(i,k+1)}\vert}{\sqrt{(b_{j(i,k)}-b_{j(i,k+1)})^2+(a_{j(i,k)}-
  a_{j(i,k+1)})^2}}
 \end{array}.
\end{equation}
In acute angles of right and obtuse triangles we have
\begin{equation} \label{eq6}
h_{i,j(i,k)}=\min\limits_k\left[L_{i,j(i,k)},L_{i,j(i,k+1)}\right].
\end{equation}
The symbol \(r_{max\,j(i,k)}\) represents a~maximal radius of the \(j\)-th point
being a~neighbour of the \(i\)-th point.
\begin{figure}[ht]
 {\centering\resizebox*{0.7\textwidth}{!}{\includegraphics{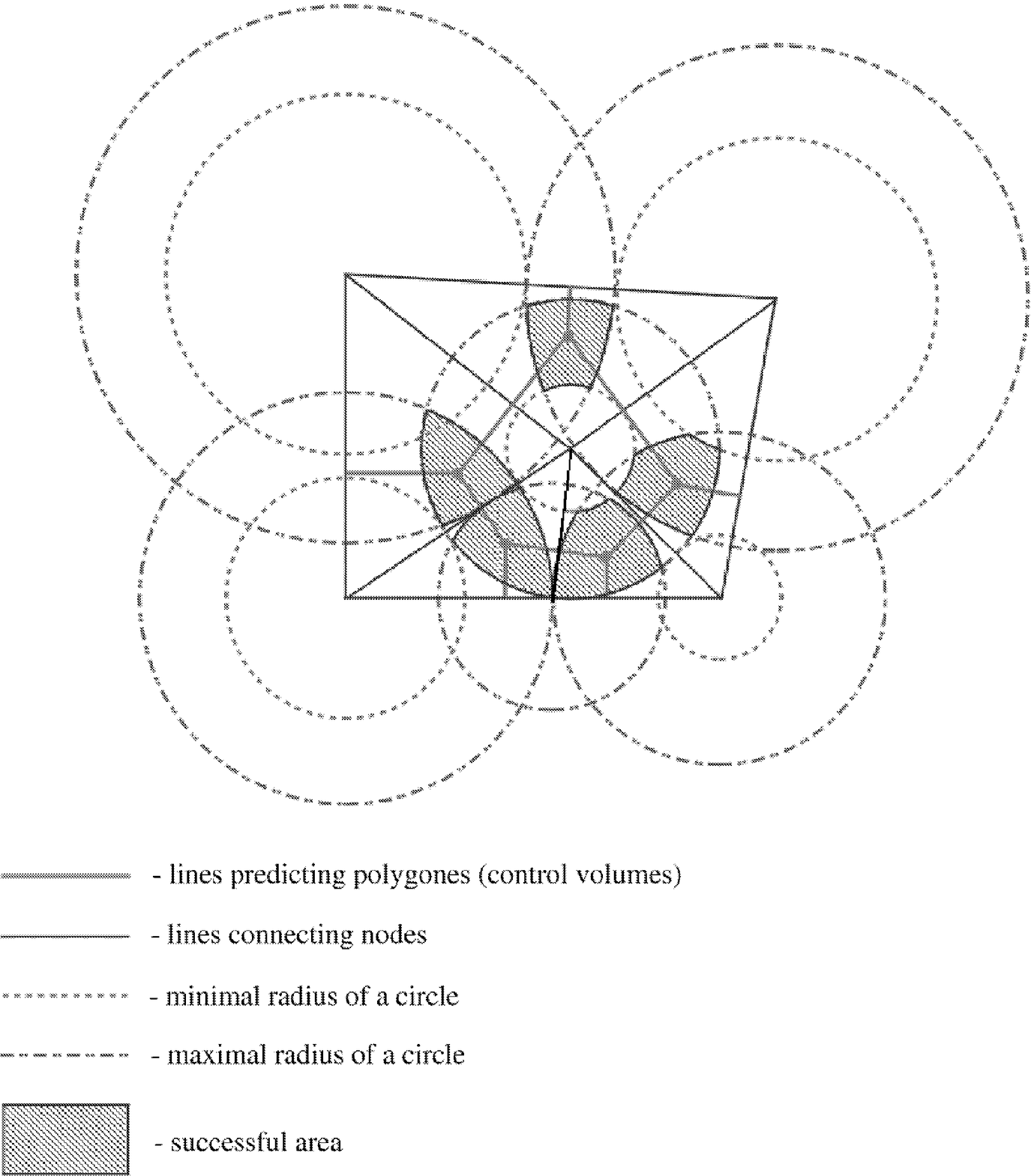}}\par}
 \caption{\label{fig2} Idea of generation of control volumes.}
\end{figure}
Using extended formula~(\ref{eq2}) and additional
expressions~(\ref{eq3})$\div$(\ref{eq6}) one can construct an unstructured mesh.
Fig.~\ref{fig2} shows an idea of the control volume generation using the
expression~(\ref{eq2}) and the limit~(\ref{eq3}). We can observe that the local
condition - an edge of the polygon has to be perpendicular to the line connecting
two points - is fulfilled. The general algorithm uses some optimisation that to
find the points~(\ref{eq2}) being the vertices of the polygons (control volumes).
In this paper we neglect the construction of an objective function because it is
not the subject of our consideration. We applied Rosenbrock's method~\cite{Kunzi}
for the local solution. For the global solution we use the soft selection
method~\cite{Galar} which is a~global optimisation method. Prediction of the
unstructured mesh under such way is not only the one way of the mesh construction.
It is possible to construct another type of mesh without restriction that
expression~(\ref{eq2}) is a~solution of eqn.~(\ref{eq1}). But we have to restrict
the expression~(\ref{eq3}) because it guaranties right mesh construction for the
control volume method~\cite{Patankar}. Let us assume that a~radius of the circle
overlaps other radiuses of the neighbour circles. Of course we have to restrict
the condition~(\ref{eq3}). Corresponding to previous and present mesh construction
we have
\begin{equation} \label{eq8}
 L_{i,j(i,k)}-\big(r_{i}+r_{j(i,k)}\big)\leq 0.
\end{equation}
Following eqn.~(\ref{eq8}) the point coordinates given by expression~(\ref{eq2})
are not the solution of the system (\ref{eq1}) but they become the coordinates of
the point predicted by lines connecting two points of crossing circles. From the
other hand, we can assume that a~radius of the circle does not overlap other
radiuses of the neighbour circles. In such situation we have
\begin{equation} \label{eq9}
 L_{i,j(i,k)}-\big(r_{i}+r_{j(i,k)}\big)>0.
\end{equation}
One can say, that conditions~(\ref{eq8}) and~(\ref{eq9}) allow us to obtain
different forms of unstructured meshes. However, in local solution is possible
some combination of expressions~(\ref{eq8}) and~(\ref{eq9}) respectively.

Similar to previous results we consider a~set of random points \(P_i(a,b,c)\) in
\(R^3\) space. Analogously to the previous considerations, we use spatial
Delaunay triangulation that to find neighbourhood of point \(P_i\) which is
defined by several points \(P_{j(i,k,l)}\). The index \(j(i,k,l)\) is a~function
which establishes a~point number having neighbourhood with the \(i\)-th point.
The temporal index \(k\) describes a~number of tetrahedrons having the common
\(i\)-th point. It varies from \(1\) to \(N(i)\). The next index \(l\) varies
from \(1\) to \(3\) and it is a~point number corresponding to the \(k\)-th
tetrahedron. The four points \(P_i,P_{j(i,k,1)},P_{j(i,k,2)},P_{j(i,k,3)}\)
define a~\(k\)-th tetrahedron. We search radiuses of spheres which centers are
located in the points. In each tetrahedron crosses the four radiuses and they
create a~point which is the vertice of the control volume. Solving the following
system
\begin{equation} \label{eq10}
 (\widehat{x}-a_{l})^{2}+(\widehat{y}-b_{l})^{2}+
 (\widehat{z}-c_{l})^{2}=r_{l}^{2}, (l=1..4),
\end{equation}
we can find the local vertices coordinates
\begin{equation} \label{eq11}
 \widehat{x}(r_{1},\, r_{2},\, r_{3},\, r_{4})=\frac{W_{x}}{W}, \hspace{2mm}
 \widehat{y}(r_{1},\, r_{2},\, r_{3},\, r_{4})=\frac{W_{y}}{W}, \hspace{2mm}
 \widehat{z}(r_{1},\, r_{2},\, r_{3},\, r_{4})=\frac{W_{z}}{W},
\end{equation}
where
\begin{eqnarray*}
 W=
 \left|
 \begin{array}{ccc}
  2(a_1-a_2) & 2(b_1-b_2) & 2(c_1-c_2) \\
  2(a_1-a_3) & 2(b_1-b_3) & 2(c_1-c_3) \\
  2(a_1-a_4) & 2(b_1-b_4) & 2(c_1-c_4) \\
 \end{array} \right|, \hspace{3mm}
 W_x=
 \left|
 \begin{array}{ccc}
  \triangle_1 & 2(b_1-b_2) & 2(c_1-c_2) \\
  \triangle_2 & 2(b_1-b_3) & 2(c_1-c_3) \\
  \triangle_3 & 2(b_1-b_4) & 2(c_1-c_4) \\
 \end{array} \right|,
\end{eqnarray*}
\begin{eqnarray*}
 W_y=
 \left|
 \begin{array}{ccc}
  2(a_1-a_2) & \triangle_1 & 2(c_1-c_2) \\
  2(a_1-a_3) & \triangle_2 & 2(c_1-c_3) \\
  2(a_1-a_4) & \triangle_3 & 2(c_1-c_4) \\
 \end{array} \right|, \hspace{3mm}
 W_z=
 \left|
 \begin{array}{ccc}
  2(a_1-a_2) & 2(b_1-b_2) & \triangle_1 \\
  2(a_1-a_3) & 2(b_1-b_3) & \triangle_2 \\
  2(a_1-a_4) & 2(b_1-b_4) & \triangle_3 \\
 \end{array} \right|
\end{eqnarray*}
and
\begin{eqnarray*}
 \triangle_1=r_2^2-r_1^2+a_1^2-a_2^2+b_1^2-b_2^2+c_1^2-c_2^2, \\
 \triangle_2=r_3^2-r_1^2+a_1^2-a_3^2+b_1^2-b_3^2+c_1^2-c_3^2, \\
 \triangle_3=r_4^2-r_1^2+a_1^2-a_4^2+b_1^2-b_4^2+c_1^2-c_4^2.
\end{eqnarray*}
Similar to previous mesh construction, we define lower and upper limits of the
radiuses variation
\begin{equation} \label{eq12}
 \max\limits_{k,l}\left[(h_{i,j(i,k,l)}-r_{max\,j(i,k)})\right]<r_{i}<
 r_{max\,i}.
\end{equation}
The symbol \(h_{i,j(i,k,l)}\) represents height of a~triangle being the wall of
the tetrahedron and for the acute triangles is defined by the following formula
\begin{equation} \label{eq121}
\begin{array}{c}
 h_{i,j(i,k,l)}=\sqrt{\frac{
   \left(BB_{12}\cdot CC_{32}-BB_{32}\cdot CC_{12}\right)^2
   \left(AA_{32}\cdot CC_{12}-AA_{12}\cdot CC_{32}\right)^2+}
   {AA_{32}^2+BB_{32}^2+CC_{32}^2}}+\\
   \sqrt{\frac{
    \left(AA_{12}\cdot BB_{32}-AA_{32}\cdot BB_{12}\right)^2}
    {AA_{32}^2+BB_{32}^2+CC_{32}^2}}
\end{array},
\end{equation}
where
\begin{eqnarray*}
 AA_{12}&=&a_i-a_{j(i,k,l)}, \hspace{5mm} AA_{32}=a_{j(i,k,l+1)}-a_{j(i,k,l)}, \\
 BB_{12}&=&b_i-b_{j(i,k,l)}, \hspace{5mm} BB_{32}=b_{j(i,k,l+1)}-b_{j(i,k,l)}, \\
 CC_{12}&=&c_i-c_{j(i,k,l)}, \hspace{5mm} CC_{32}=c_{j(i,k,l+1)}-c_{j(i,k,l)}.
\end{eqnarray*}
In acute angles of right and obtuse spatial triangles we have
\begin{equation}\label{eq122}
 h_{i,j(i,k,l)}=\min\limits_l\left[L_{i,j(i,k,l)},L_{i,j(i,k,l+1)}\right],
\end{equation}
where \(L_{i,j(i,k,l)}\) is a~spatial distance defined between two points as
\begin{equation}\label{eq123}
 L_{i,j(i,k,l)}=\sqrt{\left(a_i-a_{j(i,k,l)}\right)^2+
                      \left(b_i-b_{j(i,k,l)}\right)^2+
                      \left(c_i-c_{j(i,k,l)}\right)^2}.
\end{equation}
When the temporal index \(l+1\) exceeds \(3\) then we put
\(P_{j(i,k,l+1)}=P_{j(i,k,1)}\). Maximal radius of a~sphere is defined as
\begin{equation}\label{eq124}
 r_{max\, i}=\min\limits_k\left[H_{i,j(i,k)}\right].
\end{equation}
However, the function \(H_{i,j(i,k)}\) is height of a~tetrahedron created by the
vertices \(P_i,P_{j(i,k,1)},P_{j(i,k,2)},P_{j(i,k,3)}\) and for the acute
tetrahedron we have
\begin{equation} \label{eq13}
 H_{i,j(i,k)}=
 \frac{\vert A\cdot a_i+B\cdot b_i+C\cdot c_i+D\vert}{\sqrt{A^2+B^2+C^2}}
\end{equation}
where
\begin{eqnarray*} \label{eq13a}
 A=m_{21}\cdot n_{31}-m_{31}\cdot n_{21}, && \hspace{5mm}
 B=d_{31}\cdot n_{21}-d_{21}\cdot n_{31}, \\
 C=d_{21}\cdot m_{31}-d_{31}\cdot m_{21}, && \hspace{5mm}
 D=-A\cdot a_{j(i,k,1)}-B\cdot b_{j(i,k,1)}-C\cdot c_{j(i,k,1)}
\end{eqnarray*}
and
\begin{eqnarray*} \label{eq13b}
 d_{21}=a_{j(i,k,2)}-a_{j(i,k,1)}, \; m_{21}=b_{j(i,k,2)}-b_{j(i,k,1)}, \;
 n_{21}=c_{j(i,k,2)}-c_{j(i,k,1)}, \\
 d_{31}=a_{j(i,k,3)}-a_{j(i,k,1)}, \; m_{31}=b_{j(i,k,3)}-b_{j(i,k,1)}, \;
 n_{31}=c_{j(i,k,3)}-c_{j(i,k,1)}.
\end{eqnarray*}
But for the right and obtuse tetrahedrons we have
\begin{equation} \label{eq14}
 H_{i,j(i,k)}=\min\limits_l\left[h_{i,j(i,k,l)}\right].
\end{equation}
The symbol \(r_{max\, j(i,k)}\) represents a~maximal radius of \(j\)-th point
being a~neighbour of the \(i\)-th point.

\section{Example of calculations}
\begin{figure}[htp]
 {\centering\resizebox*{0.9\textwidth}{!}{\includegraphics{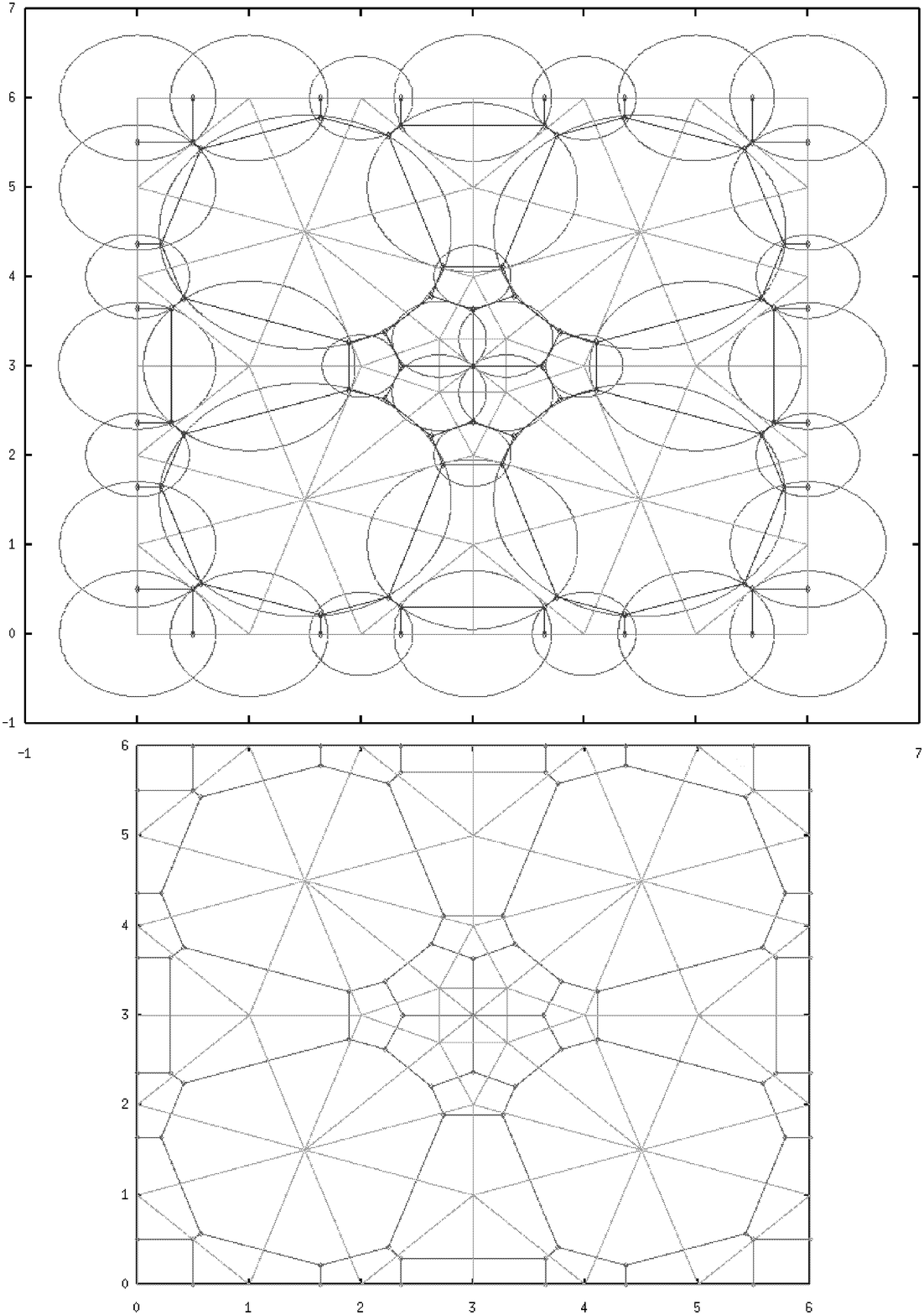}}\par}
 \caption{\label{fig4} Example of control volumes taking into consideration that
                       formula~(\ref{eq2}) is a~solution of the
                       system~(\ref{eq1}):
   		       a) random points after triangulation;
		       b) forms of control volumes.}
\end{figure}

In practical mesh construction when expression (\ref{eq3}) is fulfilled we obtain
a~unique solution. Fig.~\ref{fig4} shows an example of mesh created under
restriction that formula~(\ref{eq2}) is a~solution of the system (\ref{eq1}).
Fig.~\ref{fig4}a presents triangulation of some points in which generated
circles. Within the triangle crosses the circles only in one point. According to
the previous description we draw control volumes presented by Fig.~\ref{fig4}b.
Nevertheless, we can construct a~mesh without restriction that
formula~(\ref{eq2}) is a~solution of the system~(\ref{eq1}). We can apply
conditions~(\ref{eq8}) or~(\ref{eq9}) respectively. In this case we generate
a~mesh much more easier than in previous case. Expressions~(\ref{eq8})
and~(\ref{eq9}) extend our considerations for different generation of meshes.

\section{Concluding remarks}
In this paper we elaborate a~new method of generation of unstructured meshes. The
algorithm is prepared for the control volume method in two- and three dimensional
space. For random location of points we use Delaunay triangulation before the mesh
generation. Our method bases in 3D on circles generation in the points. Within one
triangle created by three points crosses the circles in one point. The point is
a~vertice of a~control volume. We extend our considerations for the triangles in
which are overlapping or non-overlapping circles. The extension allow us to
generate meshes much more easier than in previous case. In opposite to the Thyssen
polygons we can generate control volumes for the right and obtuse triangles.
However, our method is suitable and easy to use in generation of control volumes
in three dimensional space.

\end{document}